\documentclass[aps,pra,epsfigure,twocolumn,superscriptaddress]{revtex4-1}
\usepackage[colorlinks=true,linkcolor=blue,urlcolor=blue,citecolor=blue,pdfusetitle]{hyperref}
\usepackage[utf8]{inputenc}
\usepackage[english]{babel}
\usepackage{amsmath}
\usepackage[caption = false]{subfig}
\usepackage{graphicx,epstopdf}
\usepackage{blindtext}
\usepackage[table,xcdraw]{xcolor}
\usepackage{lipsum}
\usepackage{amsfonts}
\usepackage{bbm}
\usepackage{amssymb}
\usepackage{enumerate}
\usepackage{color}
\usepackage{latexsym}
\usepackage{physics}
\usepackage{times,txfonts}

\newcommand{\Fcal}{\mathcal{F}}

\newcommand{\Ical}{\mathcal{I}}

\newcommand{\1}{\mathbbm{1}}

\newcommand{\SubFig}[2]{\ref{#1}{\color{blue}#2}}

\definecolor{MyGreen}{RGB}{0, 179, 134}
\definecolor{MyRed}{RGB}{255, 102, 102}

\newcommand{\UFSCar}{Departamento de Física, Universidade Federal de São Carlos, \\Rodovia Washington Luís, km 235 - SP-310, 13565-905 São Carlos, SP, Brazil}
\newcommand{\SU}{Department of Physics, Stockholm University, AlbaNova University Center 106 91 Stockholm, Sweden}
\newcommand{\Nice}{Universit\'e C\^ote d'Azur, CNRS, Institut de Physique de Nice, 06560 Valbonne, France}

\newcommand{\UAM}{Depto. de F\'isica de Materiales, Instituto Nicol\'as Cabrera, Instituto de F\'isica de la Materia Condensada, Universidad Aut\'onoma de Madrid, 28049 Madrid, Spain}

\newcommand{\IPCO}{Institute of Physics, Carl von Ossietzky University, Oldenburg, 26129 Germany}

\usepackage{orcidlink}

\begin{document}

\title{Multipartite entanglement in the photon number basis by \\sequential excitation of a three-level system}

\author{Alan C. Santos~\orcidlink{0000-0002-6989-7958}}
\email{ac\_santos@df.ufscar.br}
\affiliation{\UFSCar}
\affiliation{\SU}

\author{Christian Schneider}
\affiliation{\IPCO}

\author{R. Bachelard~\orcidlink{0000-0002-6026-509X}} 
\affiliation{\UFSCar}
\affiliation{\Nice}

\author{Ana Predojevi\'c} 
\affiliation{\SU}

\author{Carlos Ant\'on-Solanas} 
\email{carlos.anton@uam.es}
\affiliation{\UAM}


\begin{abstract}
We propose a general scheme to generate entanglement encoded in the photon number basis, via a sequential resonant two-photon excitation of a three-level system. We apply it to the specific case of a quantum dot three-level system, which can emit a photon pair through a biexciton-exciton cascade. The state generated in our scheme constitutes a tool for secure communication, as the multipartite correlations present in the produced state may provide an enhanced rate of secret communication with respect to a perfect GHZ state.  
\end{abstract}

\maketitle

\begin{figure}[t!]
	\centering
	\includegraphics[width=\linewidth]{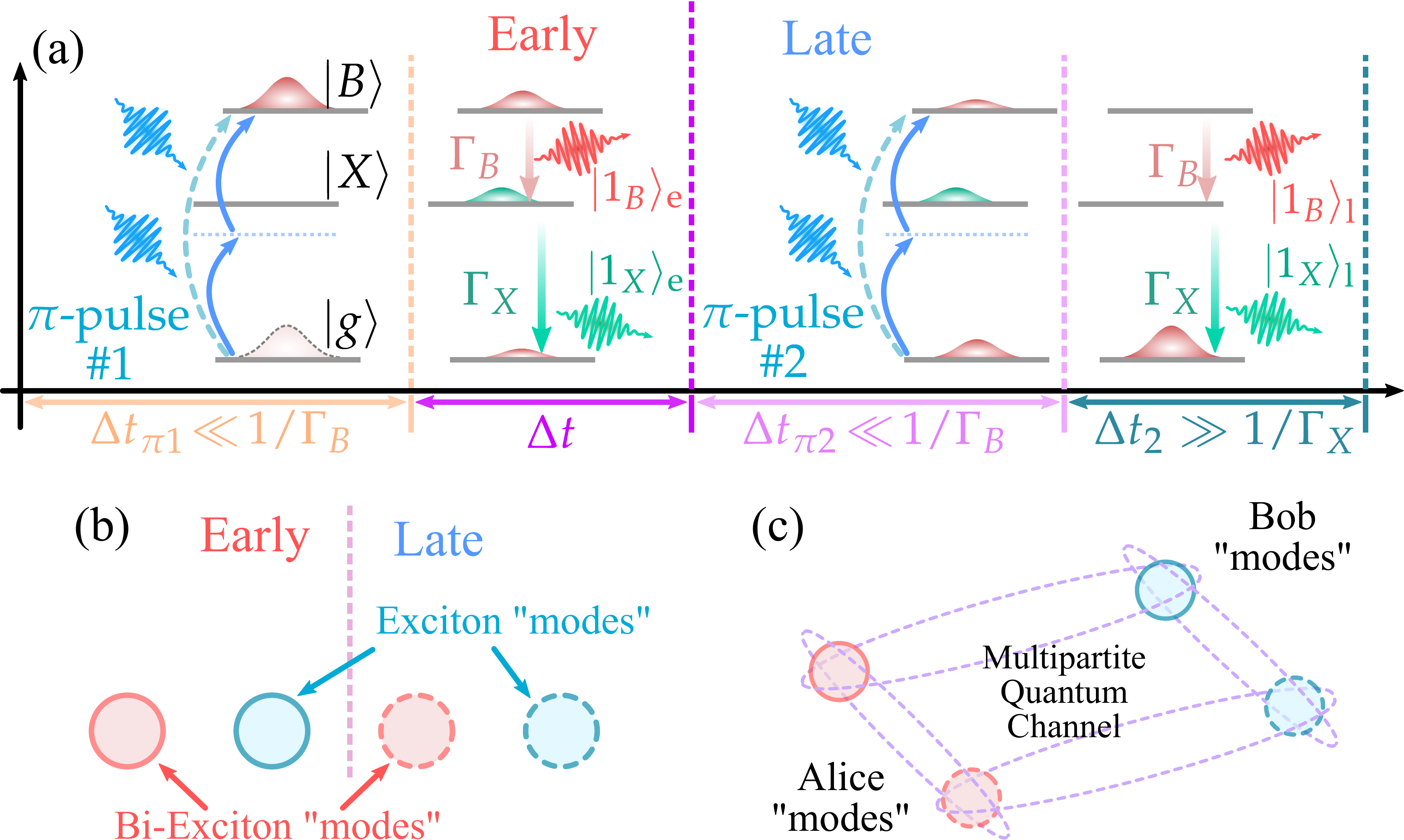}
	\caption{(a) Schematics of a three-level system excited resonantly by sequential $\pi$-pulses at the transition $\ket{g} \rightarrow \ket{B}$. The time interval between the pulses is $\Delta t$. (b) The photons emitted in this process can be classified as early and late photons depending if the emission takes place before or after the second excitation pulse. The photon pairs emitted have the energy of the biexciton/exciton responsible for their emission. Four modes are required to describe the energy and photon number subspaces. (c) The state generated is used as a multipartite quantum channel between Alice and Bob.
	}
	\label{Fig:PresentationScheme}
\end{figure}

The generation of multipartite entangled states is a challenging task which is crucial for a number of applications in quantum information processing including teleportation-based protocols~\cite{Gao:10,Ma:12,Ralph:99,Bartlett:03,Huang:04}, measurement-based quantum computation~\cite{Kiesel:05,Park:07,Kalasuwan:10,Schwartz:16}, and quantum communication~\cite{Vaziri:02,Lance:04}. {In this context, Wein et al.~\cite{Wein:22} recently proposed a deterministic scheme to generate time-entanglement in the photon-number basis, via the resonant drive of a two-level system with two delayed, instantaneous $\pi$-pulses. In a similar scheme with natural atoms, same type of temporal entanglement has been demonstrated by \'Alvarez \textit{et al}~\cite{Alvarez:22}. In the present work, we extend the simple two-pulse protocol of Ref.~\cite{Wein:22} to a three-level system (3LS)}, where these two pulses coherently drive the transition between the ground ($\ket{g}$) and the top 3LS level (labelled as $\ket{B}$, biexciton), see Fig.~\SubFig{Fig:PresentationScheme}{a}. The specific system considered here is a (three-level) single semiconductor quantum dot subject to a biexciton-exciton cascade decay.

Semiconductor quantum dots are used to generate single photons~\cite{he2013demand,tomm2021bright,somaschi2016near,Unsleber:16}, correlated photon pairs~\cite{PhysRevB.66.045308}, cluster states~\cite{Schwartz:16,2022arXiv220709881C,cogan2023deterministic} and entangled photon pairs~\cite{PhysRevLett.96.130501,jayakumar2014time,PhysRevLett.121.110503}. Recently, the coherent driving of a two-level system quantum dot (a charged exciton under zero magnetic field) has allowed to {demonstrate the coherent superposition~\cite{loredo2019,Karli:23,Maillette:22} and time-entanglement~\cite{Wein:22,Alvarez:22} encoded in the photon-number basis, a basis that has been scarcely exploited due to the difficulty to manipulate photonic states containing vacuum~\cite{PhysRevLett.88.070402,Alvarez_2022,PhysRevA.66.032307}}.
In a quantum dot system, the biexciton decays to the first excited level ($\ket{X}$, exciton) by emitting a photon $\ket{1_{B}}$, at a decay rate $\Gamma_{B}$. Then the system spontaneously decays from the exciton level, at a rate $\Gamma_X$, to the ground state $\ket{g}$, emitting a photon in the state $\ket{1_{X}}$. The delay $\Delta t$ between the two pulses must be on the order of $1/\Gamma_{X}$, since the second pulse must re-excite the three-level system before the spontaneous emission of the cascade is finished. We identify two time windows, "early" (e) and "late" (l), where "early" corresponds to the time interval between the two pulses ($\Delta t$), and "late" is the time window denoting the time elapsed between the second pulse and the total decay of the three-level system to the ground state (see time axis in Fig.~\SubFig{Fig:PresentationScheme}{a}). In this simplified 3LS picture, we consider a single biexciton-exciton cascade, correctly selecting one of the two decaying polarisations of the full cascade; this way the potential exciton fine-structure splitting does not affect the final state along the decay process.


The system considered in this work can be described by a four-mode system, as illustrated in Fig.~\SubFig{Fig:PresentationScheme}{b}. Considering each mode as a two-level system, we show how to control the biexciton-exciton decay cascade to generate multipartite entanglement between such modes. Furthemore, we will show how such a system is useful for quantum communication tasks, as provided by the analysis of the optimal rate of secret communication~\cite{Sharma:20} associated to such a state (Fig.~\SubFig{Fig:PresentationScheme}{c}). 


\begin{figure}[t!]
\centering
\includegraphics[width=\linewidth]{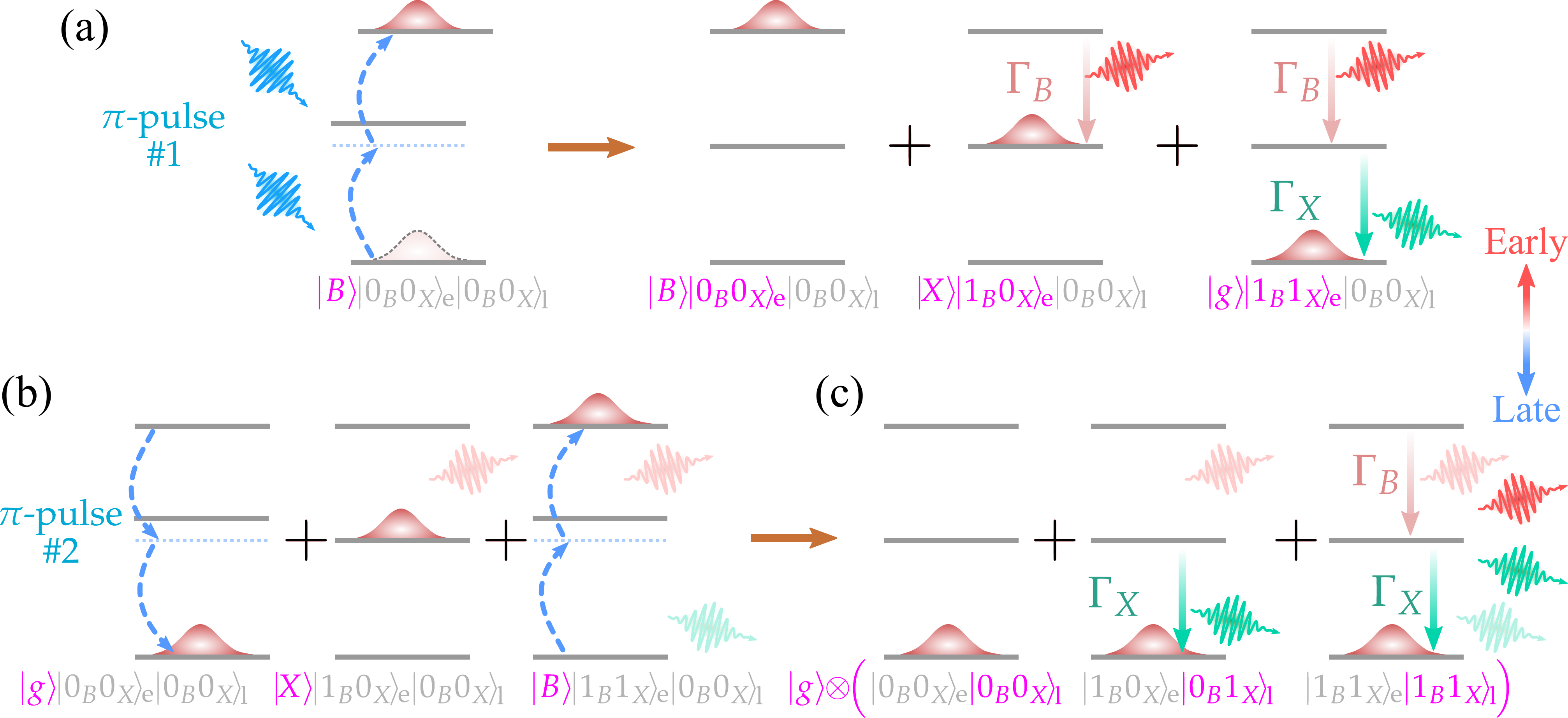}
\caption{(a) Schematic representation of the decay dynamics during the early time window, when the three-level system is initially excited to the biexciton state. The system composed of the dot plus the photonic modes can decay through three possible states involving only early photons. (b) After the delay time $\Delta t$ (first decay process), the second pulse is sent, which only affects the states $\ket{g}$ and $\ket{B}$. (c) Finally, the system decays in the ground state, emitting late photons.}
\label{Fig:SchemeState}
\end{figure}

Apart from the intrinsic limitation to the photon indistinguishability in the biexciton-exciton cascade determined by the
biexciton and exciton decay rates~\cite{PhysRevLett.125.233605} (which still could be engineered via photonic cavity effects), our model considers that the cascade is free from other additional dephasing mechanisms (such as phonons, or electrons; photons, although
spontaneously emitted, are described in the total state). The multipartite entangled state is generated through the following sequence: at time $t=0$, {the three-level system (initialized in the ground state) is driven with a fast $\pi$-pulse, which brings it into the biexciton level $\ket{B}$~\cite{Jayakumar:13,muller2014demand}}. Its coupling with the vacuum state $\ket{0_{B}0_{X}}_{\mathrm{e}}$ of the "early" mode induces the system to spontaneously decay towards the ground state. After a time $\Delta t$ the state of the system reads $\ket{\Psi(\Delta t)} = \ket{\psi(\Delta t)}_{\mathrm{e}} \ket{0_{B}0_{X}}_{\mathrm{l}}$, where $\ket{\psi(\Delta t)}_{\mathrm{e}}$ is a superposition of three possible states in the early time window, as shown in Fig.~\SubFig{Fig:SchemeState}{a}. These three possible states are: i) the three-level system remains in the biexciton level, ii) it emits an early biexciton photon and ends up in the exciton state $\ket{X}$, or iii) the system fully decays into the ground state $\ket{g}$ by emitting both early biexciton and exciton photons (completing the full decay of the cascade). Therefore, the system state can be written as
\begin{equation}
\ket{\psi(\Delta t)}_{\mathrm{e}} = \alpha \ket{B}\ket{0_{B}0_{X}}_{\mathrm{e}} + \beta \ket{X}\ket{1_{B}0_{X}}_{\mathrm{e}} +\gamma \ket{g}\ket{1_{B}1_{X}}_{\mathrm{e}} ,
\end{equation}
with $\gamma = \sqrt{1 - \alpha^2 - \beta^2 }$,
\begin{equation}
\alpha = e^{-\frac{1}{2} \Gamma _{B} \Delta t} , ~ \mathrm{and} ~ \beta = \sqrt{\Gamma _{B}\left(e^{-\Gamma _{B} \Delta t}-e^{-\Gamma _{X} \Delta t}\right)/(\Gamma _{X}-\Gamma _{B})} . 
\end{equation}

After a time $\Delta t$, a second $\pi$-pulse is applied on the transition $\ket{B}\rightleftarrows\ket{g}$, setting the time from which "late" photons can be emitted. Immediately after this pulse, the state of the system reads $\ket{\Psi_{\#2}} = \ket{\psi_{\#2}} \ket{0_{B}0_{X}}_{\mathrm{l}}$ with $\ket{\psi_{\#2}} = \alpha \ket{g}\ket{0_{B}0_{X}}_{\mathrm{e}} + \beta \ket{X}\ket{1_{B}0_{X}}_{\mathrm{e}} +\gamma \ket{B}\ket{1_{B}1_{X}}_{\mathrm{e}}$, as plotted in Fig.~\SubFig{Fig:SchemeState}{b}. The population in the exciton state does not change since the driving pulse is not resonant with this level~\cite{Jayakumar:13}. Finally, the system emits photons into the "late" mode (initially in the vacuum state $\ket{0_{B}0_{X}}_{\mathrm{l}}$), decaying to the ground state (see Fig.~\SubFig{Fig:SchemeState}{c}). The final state of the system is $\ket{\Psi_{\mathrm{fin}}} = \ket{g} \ket{\psi_{\mathrm{fin}}}$, where the photonic part corresponds to a photon-number entangled state, 
\begin{equation}
\ket{\psi_{\mathrm{fin}}} = \alpha \ket{0_{B}0_{X}}_{\mathrm{e}}\ket{0_{B}0_{X}}_{\mathrm{l}} + \beta \ket{1_{B}0_{X}}_{\mathrm{e}}\ket{0_{B}1_{X}}_{\mathrm{l}} +\gamma \ket{1_{B}1_{X}}_{\mathrm{e}}\ket{1_{B}1_{X}}_{\mathrm{l}} . \label{Eq:State}
\end{equation}

Equation \ref{Eq:State} describes a four-mode state in a linear superposition of vacuum and single photons in each mode, in this way each mode represents a two-level system with (computational basis) states $\ket{0}$ and $\ket{1}$. As a first remark, we note that due to the finite probability $\beta$ of decaying into the state $\ket{1_{B}0_{X}}_{\mathrm{e}}\ket{0_{B}1_{X}}_{\mathrm{l}}$, $\ket{\psi_{\mathrm{fin}}}$ is never a maximally entangled four-mode state, like the GHZ state $\ket{\mathrm{GHZ}}=(\ket{0000}+ \ket{1111})/\sqrt{2}$. In fact, by imposing $\alpha = 1/\sqrt{2}$ we get the condition $\Gamma _{B} \Delta t = \log 2$, which leads to a vanishing $\beta$ only in the limit $\Gamma _{X}/\Gamma _{B} \gg 1$. In addition, the regime of parameters where the fidelity of $\ket{\psi_{\mathrm{fin}}}$ is larger than that of a GHZ state, $\Fcal = |\braket{\mathrm{GHZ}}{\psi_{\mathrm{fin}}}|^2$,corresponds to a cascade decay $\Gamma_{X} > \Gamma_{B}$. However, it does not mean that state (\ref{Eq:State}) is not entangled when $\Gamma_{X} \leq \Gamma_{B}$. 

\begin{figure}[t!]
\centering
\includegraphics[width=\linewidth]{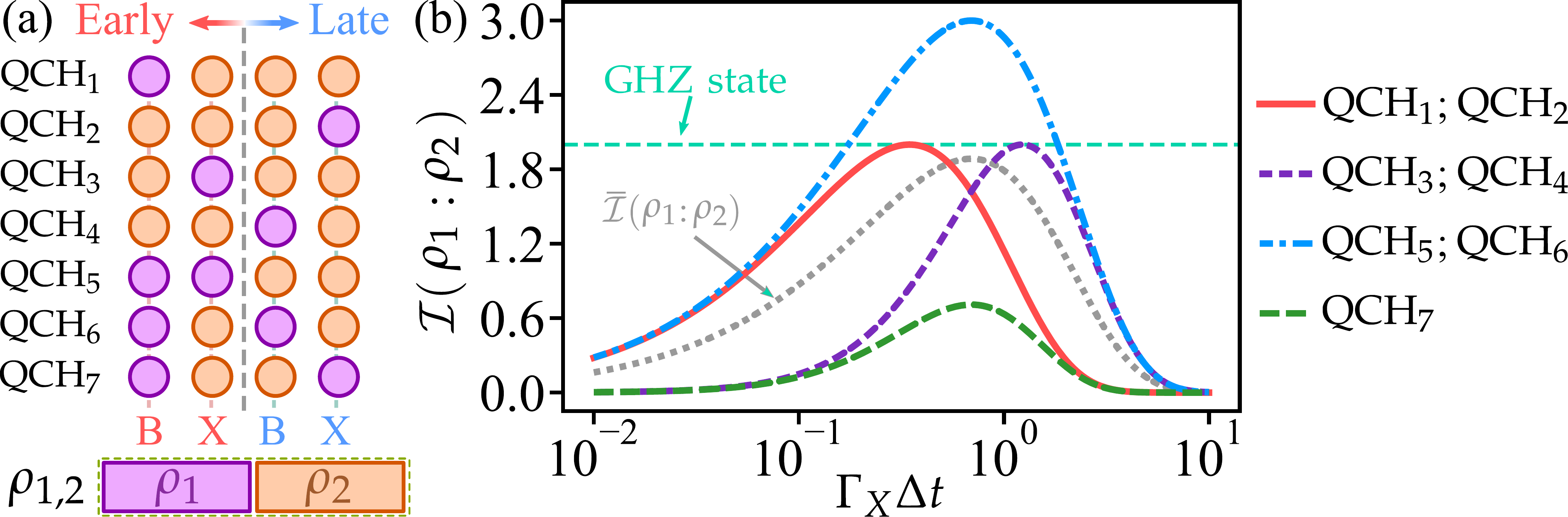}
\caption{(a) Available quantum channels associated to state (\ref{Eq:State}), where purple and orange modes denote subspace $P_{1}$ and $P_{2}$, respectively, for each channel. biexciton (B) and exciton (X) "modes" for Early and Late time windows are ordered as (from left to right): Early-B, Early-X, Late-B, Late-X. (b) Mutual information for each channel available in our system, where the average mutual information over all possibilities, $\bar{\Ical}(\rho_{1}:\rho_{2})$, shows that a GHZ state has stronger correlations than state (\ref{Eq:State}).}
\label{Fig:Correlations}
\end{figure}

We quantify the multipartite entanglement using mutual information between two partitions $P_{1}$ and $P_{2}$, defined as
\begin{align}
\Ical(\rho_{1}:\rho_{2}) = S(\rho_{1}) + S(\rho_{2}) - S(\rho_{1,2}) ,
\end{align}
where $S(\rho) = - \mathrm{tr}\left[\rho\log_{2}(\rho)\right]$ is the von Neumann entropy of $\rho$. Here, $\rho_{1,2}$ is the density matrix of the whole system, while $\rho_{1}$ and $\rho_{2}$ are the reduced density matrices of the subspaces $P_{1}$ and $P_{2}$, respectively. Because we consider pure states in this work, one finds $S(\rho_{1,2}) = 0$, which means that $\Ical(\rho_{1}:\rho_{2}) >0$ indicates entanglement between the system $P_{1}$ and $P_{2}$. Given the set of four two-level systems used to encode information and using $\Ical(\rho_{1}:\rho_{2}) = \Ical(\rho_{2}:\rho_{1})$, the different possible decompositions of the system state along subsystems $P_1$ and $P_2$ are shown in Fig.~\SubFig{Fig:Correlations}{a}. Motivated by the exploration of quantum communication applications, we refer to the $n$-th composition as the $n$-th QCH (quantum channel) available in our system. Hence, by assuming values for the decay rates given by $\Gamma_{B}/\Gamma_{X}\approx 2$~\cite{Simon:05}, we compute all bipartite correlations of state (\ref{Eq:State}) with respect to the delay time $\Delta t$ between the pulses (in the Supplementary Material~\cite{SM} we present further discussion on other regimes of the ratio $\Gamma_{B}/\Gamma_{X}$). We note that the intrinsically-limited indistinguishability of the cascade~\cite{PhysRevLett.125.233605} could have an impact on the analysis of the mutual information if additional dephasing times of biexciton and exciton are very different. We would like to indicate that the application of the protocol discussed does not involve interference between emitted photonic states. However, the stable phase between two parts of a bipartite entanglement is very important for entanglement: if the phase coherence is erased, the entanglement disappears (see details in~\cite{SM}). Also, the mutual information or the maximal population drop rapidly and becomes small when the pure	phasing rates are larger than the corresponding decay rates~\cite{SM}.


As shown in Fig.~\SubFig{Fig:Correlations}{b}, the quantum correlations are evenly distributed over the system for all bipartite choices of a four-qubit GHZ state. In other words, $\Ical_{\mathrm{GHZ}}(\rho_{1}:\rho_{2})=2$ for any composition $P_{1,2} = P_{1}\otimes P_{2}$. However, state (\ref{Eq:State}) exhibits a high degree of correlations only for a specific set of subsystems $P_1$ and $P_2$, which are the ones most appropriate for the realization of quantum tasks. The results shown in Fig.~\SubFig{Fig:Correlations}{b} can be interpreted as follows: i) The amount of multipartite entanglement created between $P_{1}$ and $P_{2}$ can be controlled through the delay time $\Delta t$ between the early and late pulses. ii) The fact that $\Ical(\rho_{1}:\rho_{2})$ depends on the QCH used means that the entanglement is not spread equally over the entire system. iii) Regarding the QCHs in which $\Ical(\rho_{1}:\rho_{2})>\Ical_{\mathrm{GHZ}}(\rho_{1}:\rho_{2})$, they should not be understood as being more correlated than a GHZ state. This result rather indicates that, for such a channel, there are higher correlations between $P_{1}$ and $P_{2}$ than for its GHZ counterpart, yet the modes inside each subspace are weakly correlated. By computing the average multipartite entanglement as $\bar{\Ical}(\rho_{1}:\rho_{2})=(1/7)\sum_{n=1}^{7}\Ical_{n}(\rho_{1}:\rho_{2})$, with $\Ical_{n}(\rho_{1}:\rho_{2})$ the mutual information for the $n$-th QCH, one gets $\bar{\Ical}(\rho_{1}:\rho_{2})<\Ical_{\mathrm{GHZ}}(\rho_{1}:\rho_{2})$.

Let us now discuss how to use state (\ref{Eq:State}) for quantum communication tasks, given entanglement distribution over the system (see Fig.~\ref{Fig:Correlations}). In particular, here we focus on the secure communication between two parties (Alice and Bob) subject to the malicious interference of a third party (Eve). The scenario is as follows: Alice and Bob share a quantum channel given by one of the bi-partitions shown in Fig.~\SubFig{Fig:Correlations}{a}, in such a way that the number of modes in Bob's subspace is at least equal to Alice's one (that is QCH 1, 2, 4 or 5). We define that Eve is able to make measurements on part of Bob's system in order to get access to the information that Alice wants to share with Bob. Obviously, the channel Alice-Bob is not $100\%$ secure and the question here is: In presence of Eve's influence, what is the rate of secret communication between Alice and Bob? This question can be efficiently addressed by using the notion of \textit{conditional quantum one-time pad}~\cite{Sharma:20}, which states that the \textit{optimal} rate for secure communication is given by the conditional mutual information
\begin{equation}
\Ical(\mathrm{Alice}:\mathrm{Bob}|\mathrm{Eve}) = \Ical(\rho_{\mathrm{Alice}}:\rho_{\mathrm{BobEve}}) - \Ical(\rho_{\mathrm{Alice}}:\rho_{\mathrm{Eve}}) . \label{Eq:Rate}
\end{equation}

$\rho_{\mathrm{Alice}}$ is the reduced density matrix for Alice, $\rho_{\mathrm{BobEve}}$ the reduced state for the whole system of Bob's lab (under the control of Bob and Eve), and $\rho_{\mathrm{Eve}}$ is the reduced state for the system under Eve's control. From above equation one concludes that the greater the correlation between Alice and Bob modes, the better suited for communication the quantum channel is. For this reason, in our analysis we will focus on the channels CH$_{1}$ and CH$_{5}$. On the other hand, the greater the correlation between Alice and Eve modes, the less secure the communication between Alice and Bob will be. This analysis suggests that, for a GHZ state, Eve can affect the secure communication regardless of which of Bob's particle she has access to, as the secure communication is equally affected. 

\begin{figure}[t!]
\centering
\includegraphics[width=\linewidth]{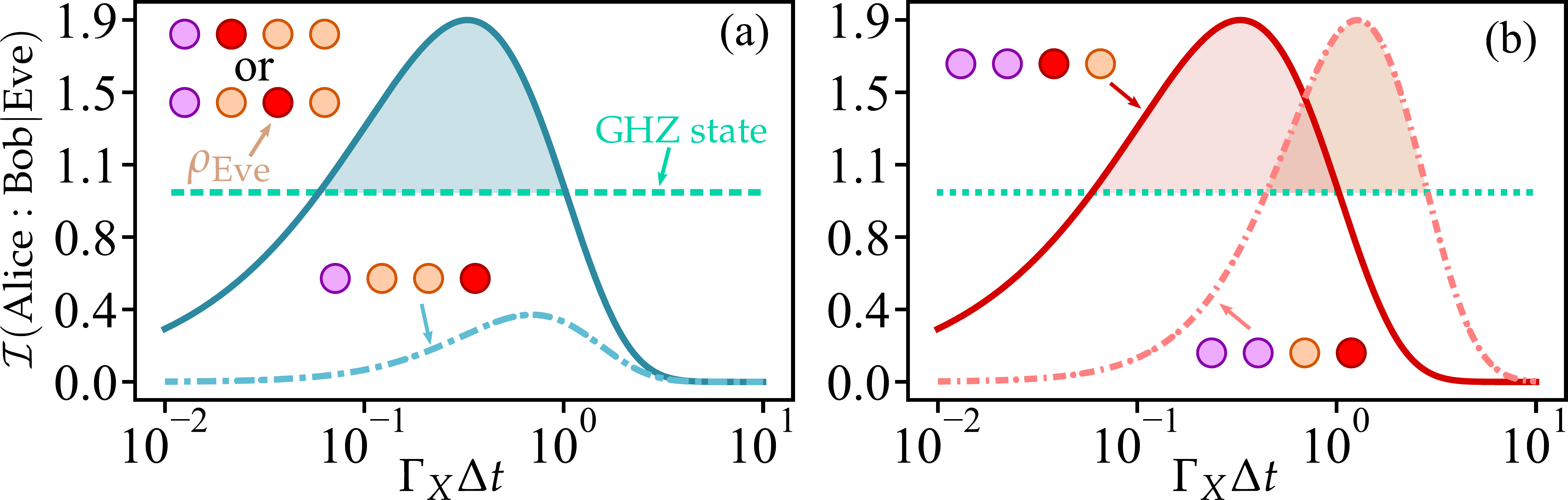}
\caption{Optimal secure rate communication between Alice and Bob, subject to Eve's interference for (a) the quantum channel CH$_{1}$ and (b) CH$_{5}$. The horizontal dashed lines represent the case where the shared state is the GHZ state. Each curve correspond to a different channel with possible Eve's influence, in which Alice and Bob subspace are denoted by purple and orange modes, respectively, while Eve has access to the red particle of the Bob's lab.}
\label{Fig:Secure}
\end{figure}

In the scheme using CH$_{1}$, Alice has access to the early biexciton, while the rest of the system is in Bob's lab. It means that Eve can, in principle, get one of the Bob's modes and perform measurements on it. As shown in Fig.~\SubFig{Fig:Secure}{a}, the correlations are not identically distributed over the system (see Fig.~\ref{Fig:Correlations}). Therefore, there is a situation where the particle under Eve's control does not have a significant impact on the secure communication, in comparison with its GHZ counterpart (horizontal dashed line). The secure rate can be increased by up to $90\%$ as compared to the GHZ state, in the appropriate time window, which corresponds to the state generated for a delay time $\Gamma_{X} \Delta t \approx 1/3$. On the other hand, if Eve has access to the correct subspace (in this case, the exciton-late subspace), the Alice-Bob quantum channel becomes much more vulnerable to Eve's attacks. In fact, for such a case there is no choice over the time $\Gamma_{X} \Delta t$ able to provide a robust channel between Alice and Bob. 
Therefore, concerning the GHZ quantum channel, the state considered in this work can be useful to \textit{probabilistically} prevent Eve's attacks, as shown in Fig.~\ref{Fig:Secure}. The probabilistic nature provided by the channel CH$_{1}$ can be bypassed if we choose channel CH$_{5}$, where Alice and Bob have the same number of modes. In this case, there is a range of values of the delay in which, regardless of Eve's subspace, the communication can be done with an optimal rate of secure communication, higher than its GHZ counterpart. For the sake of comparison, it is important to mention that, in this scenario, the optimal protocol for communication is the case where Alice and Bob share a 3 qubit GHZ state, and Eve's particle is fully uncorrelated from Alice's system, such that one gets $\Ical(\mathrm{Alice}:\mathrm{Bob}|\mathrm{Eve}) = \Ical(\rho_{\mathrm{Alice}}:\rho_{\mathrm{BobEve}})=3.0$.

In conclusion, in this work we proposed a protocol to generate multipartite entanglement in photon number modes through the decay cascade in a three-level system. Our scheme generates entangled states for an appropriate choice of the time interval between the sequential pulses $\Delta t$. 
Due to the way the correlations spread over the modes, a perfect GHZ state cannot be created. However, one can take advantage of this non-uniform distribution of correlations for secure communication. To illustrate this point, we computed the optimal rate for secure communication~\cite{Sharma:20}, in presence of an eavesdropper, considering different choices of quantum channels available in our four-mode entangled state. This allowed us to identify channels such that the state provided by our protocol provides an enhanced channel for secure communication, which is a (up to) $90\%$ improvement over the maximally entangled GHZ. 

Using the full biexciton-exciton cascade with two polarisation decay paths, where more degrees of freedom are available than for two-level systems, our scheme could lead to a more complex state than the one described in this work, and thus potentially enhanced schemes for entanglement generation and secure communication. Furthermore, as mentioned in Ref.~\cite{Wein:22}, driving the atom with $N$ consecutive pulses would be useful for increasing the temporal entanglement structure of the final photonic state. Finally, we note that our protocol exclusively considers the simplest case of two resonant pulses driving the ground-biexciton transition; however, the excitation scheme can be expanded to pulse sequences addressing different levels, generating photonic states with different entanglement characteristics. 

\textbf{Funding} ~~~ The authors thank to Dr. Stephen C. Wein for useful discussions and for carefully reading the manuscript. A.C.S. and R.B. are supported by the São Paulo Research Foundation (FAPESP) (Grants No. 2019/22685-1, 2021/10224-0, 2018/15554-5 and 2020/00725-9). 
C.A.S. Comunidad de Madrid fund ``Atracción de Talento, Mod. 1", Ref. 2020-T1/IND-19785; Grant No. PID2020113445-GB-I00 funded by the Ministerio de Ciencia e Innovación (10.13039/501100011033); the grant ULTRA-BRIGHT from the Fundación Ramon-Areces in the ``XXI Concurso Nacional para la adjudicación de Ayudas a la Investigación en Ciencias de la Vida y de la Materia"; the ``Leonardo grant to researchers in Physics 2023" from the BBVA Foundation.
A.P. acknowledges the support of the Swedish Research Council via Network grant for international researcher exchange Brazil-Sweden (SPRINT 2020-00116). 
C.S acknowledges funding within the QuantERA II Programme that has received funding from the European Union's Horizon 2020 research and innovation programme under Grant Agreement No 101017733, and with funding organisations the German ministry of education and research (BMBF) within the projects EQUAISE and TubLan Q.0. 
A.P. and R.B. also thank the Joint Brazilian-Swedish Research Collaboration (CAPES-STINT), grant 88887.304806/2018-00 and BR2018-8054. This work has been supported by the French government, through the UCA J.E.D.I. Investments in the Future project managed by the National Research Agency (ANR) with the reference number ANR-15-IDEX-01.

\textbf{Disclosures} ~~~ The authors declare no conflicts of interest.

\textbf{Statement} ~~~ See Supplement 1 for supporting content.


%

\end{document}